\documentclass[letter,twocolumn]{jpsj3}
\usepackage[usenames]{color}
\title{Separation of Energy Scales in Undoped YbRh$_2$Si$_2$ Under Hydrostatic Pressure}

\author{Yoshi \textsc{Tokiwa}$^{1}$\thanks{ytokiwa@gwdg.de},
Philipp \textsc{Gegenwart}$^{1}$, Christoph \textsc{Geibel}$^{2}$
and Frank \textsc{Steglich}$^{2}$}
\inst{$^{1}$ I. Physik. Institut, Georg-August-Universit\"{a}t G\"{o}ttingen, D-37077 G\"{o}ttingen \\
$^{2}$Max-Planck Institute for Chemical Physics of Solids, D-01187 Dresden, Germany}

\abst{The temperature ($T$)-magnetic field ($H$) phase diagram of
YbRh$_2$Si$_2$ in the vicinity of its quantum critical point is
investigated by low-$T$ magnetization measurements. Our analysis
reveals that the energy scale $T^\star(H)$, previously related to
the Kondo breakdown and terminating at 0.06~T for $T\rightarrow 0$,
remains unchanged under pressure, whereas the antiferromagnetic
critical field increases from 0.06~T ($P=0$) to 0.29~T
($P=1.28$~GPa), resulting in a crossing of $T_N(H)$ and
$T^\star(H)$. Our results are very similar to those on
Yb(Rh$_{1-x}$Co$_x$)$_2$Si$_2$, proving that the Co-induced disorder
can not be the reason for the detachment of both scales under
chemical pressure.}

\kword{Heavy fermion, Quantum critical point}
\begin{document}\maketitle
Heavy fermion (HF) metals, i.e., periodic lattices of certain
f-elements, are ideally suited to study the interplay of competing
interactions (Kondo- vs RKKY-interaction), which can lead to a
continuous phase transition at zero-temperature, driven by pressure,
doping or magnetic field~\cite{gegenwart-review}. In the approach of
the quantum critical point (QCP) the magnetic order parameter
fluctuations grow continuously in spatial and temporal dimensions,
causing strong deviations from Landau's Fermi liquid (LFL) theory.
Such non-Fermi liquid (NFL) states are characterized by a
single-excitation energy scale, which vanishes at the
QCP~\cite{hertz,millis,moriya}. This results, e.g., in a divergence
of the Gr\"uneisen ratio, $\Gamma\propto \beta/C$, or magnetic
Gr\"uneisen parameter $\Gamma_{mag}\propto (-dM/dT)/C$, ($\beta$:
volume thermal expansion, $C$: electronic specific heat, $M$:
magnetization) for QCPs tuned by pressure and magnetic field,
respectively~\cite{zhu}. Experiments on the HF metals
CeCu$_{6-x}$M$_x$ (M=Au~\cite{schroeder},Ag~\cite{Kuechler-2004})
and YbRh$_2$Si$_2$~\cite{custers,paschen,gegenwart-science} are
incompatible with the standard picture of an antiferromagnetic (AF)
QCP, which describes a spin-density-wave transition. Unconventional
quantum criticality~\cite{Si,Coleman,Senthil,Pepin},
which qualitatively differs from the predictions of the standard
theory, may arise due to a destruction of Kondo screening, leading
to a decomposition of the heavy quasiparticles into conduction
electrons and local magnetic moments.

In this letter, we focus on the clean stoichiometric HF metal
YbRh$_2$Si$_2$, located very close to a QCP~\cite{trovarelli}.
Because of the very weak AF order below $T_{\rm N}$=0.07\,K, a tiny
variation of an external control parameter is sufficient to tune the
system through the QCP. The AF order may be suppressed either by
small amounts of Ge-, La, or Ir-doping~\cite{custers,
ferstl-LaYbRh2Si2,friedemann} or magnetic fields of $H_{\rm
N}\approx 0.06$\,T and $\approx 0.7$~T, applied
perpendicular and parallel to the tetragonal $c$-axis,
respectively~\cite{gegenwart-prl02}. At $H>H_{\rm N}$, heavy LFL
behavior is found in the low-$T$ electronic specific heat, magnetic
susceptibility and electrical resistivity, with diverging
coefficients in the approach of the critical
field~\cite{gegenwart-prl05}. Correspondingly, a stronger than
logarithmic divergence of $C(T)/T$ and linear $T$-dependence of the
electrical resistivity has been found in the quantum critical
regime~\cite{custers}. These NFL effects, together with the
Gr\"uneisen ratio divergences~\cite{Kuechler,tokiwa-prl09} could not
be described within the itinerant theory for an AF QCP. Furthermore,
evidence for strong ferromagnetic (FM) fluctuations, competing with
the AF ones near the QCP have been found in bulk
suscpetibility~\cite{gegenwart-prl05} and nuclear magnetic-resonance
experiments~\cite{ishida-prl02}. A quantum tricritical-point scenario has been proposed
to account for these observations~\cite{Imada}. However, this model
predicts a saturation of the specific heat coefficient in the
approach of the QCP, as in the standard theory~\cite{moriya},
in contrast to the specific heat results. The absence of
a metamagnetic signature in the magnetization at the field-tuned
QCP~\cite{gegenwart-prl02} seems also incompatible with a
field-driven critical valence transition~\cite{watanabe}.
Alternatively, the quasiparticle mass divergence may hint at a
destruction of the Kondo effect~\cite{custers}. Indeed, a drastic
change of the Hall coefficient upon tuning through a line
$T^\star(H)$ which vanishes at $H=H_{\rm N}$ at the QCP has been
found, suggesting a strong change of the Fermi volume due to the
localization of the 4f-electrons at the QCP~\cite{paschen}.
Thermodynamic and transport measurements such as magnetization,
magnetostriction, magnetoresistance all revealed related crossovers
whose full widths at half maxima (FWHM) vanish in the $T\rightarrow
0$ limit~\cite{gegenwart-science}. Further thermodynamic evidence
for the existence of this additional energy scale arises from maxima
in the field dependence of $-\Delta M/\Delta T\approx -dS/dH$,
indicating a characteristic reduction of spin entropy upon crossing
$H^\star$, i.e., the field corresponding to $T^\star(H)$, at
constant temperature~\cite{tokiwa-prl09}.

To study the interplay of the Kondo-breakdown with the AF QCP in
YbRh$_2$Si$_2$, slightly doped single crystals in which Rh has been
partially substituted by isoelectronic but either smaller Co or
larger Ir have recently been investigated at low
temperatures~\cite{friedemann}. Most remarkably, the scale
$T^\star(H)$ remains virtually unchanged when the boundary of the AF
ground state $T_{\rm N}(H)$ is either enlarged or suppressed in the
former and latter case, respectively. For
Yb(Rh$_{1-x}$Ir$_x$)$_2$Si$_2$, a novel spin-liquid type ground
state emerges, in which the f-moments are neither Kondo-screened nor
ordered, whereas for Yb(Rh$_{1-x}$Co$_x$)$_2$Si$_2$ the field-tuned
QCP at $H_{\rm N}$ may be of itinerant nature, since it is located
within the large Fermi-surface regime at
$T<T^\star(H)$~\cite{friedemann}. Since disorder may strongly
influence quantum criticality in HF systems, we have performed
hydrostatic pressure experiments on clean undoped YbRh$_2$Si$_2$ and
compare the results with ambient-pressure ones on
Yb(Rh$_{1-x}$Co$_x$)$_2$Si$_2$.

\begin{figure}[tb]
\begin{center}
\vspace{0.5cm}
\includegraphics[width=\linewidth,keepaspectratio]{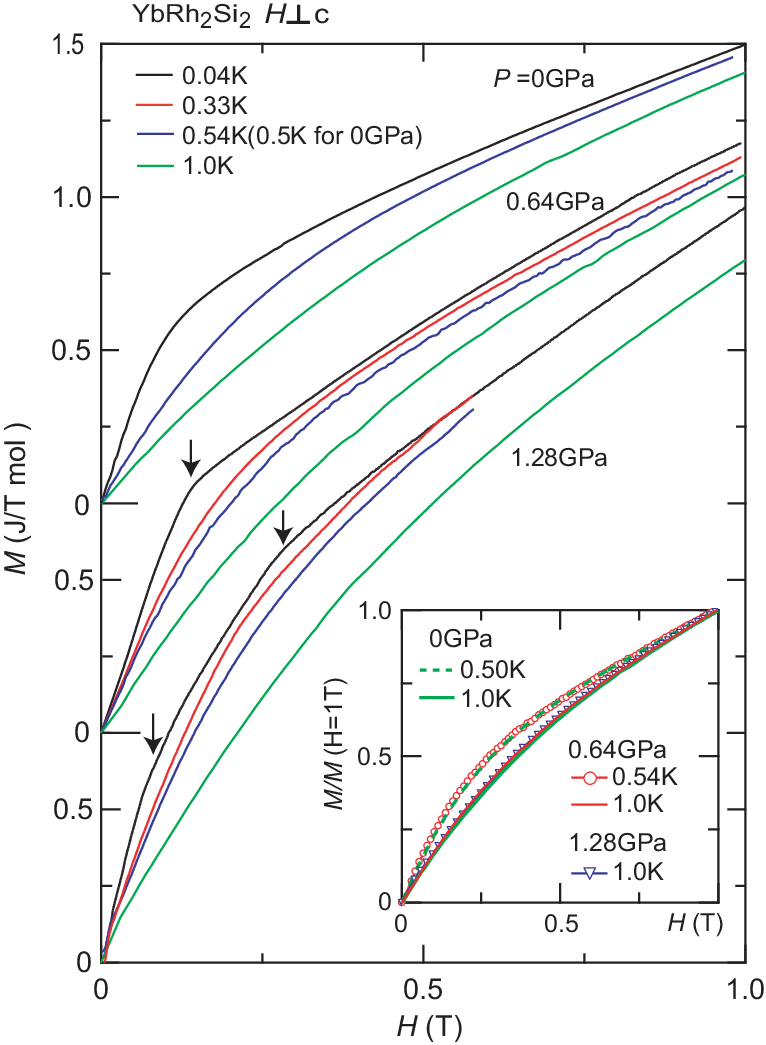}
\end{center}
\caption{(Color online) Magnetization curves of YbRh$_2$Si$_2$ in a field
perpendicular to c-axis at pressures 0, 0.64 and 1.28\,GPa for
different temperatures. Curves for 0 and 0.64\,GPa are shifted
vertically for clarity. Arrows indicate AF phase transitions. Inset:
magnetization, normalized by its value at 1\,T, vs field.}\label{f1}
\end{figure}

High-quality single crystals ($\rho_0=1~\mu\Omega$cm) were grown
from In-flux as described earlier~\cite{trovarelli}. The DC
magnetization was measured utilizing a high-resolution capacitive
Faraday magnetometer~\cite{sakakibara-faraday}. In order to
determine the magnetization under hydrostatic pressure, a
miniaturized CuBe piston-cylinder pressure cell of 6\,mm outer
diameter and 3.2\,g total weight has been designed. The piston is
made from NiCrAl, a hard material with a relatively small
magnetization. The magnetization of the pressure cell including the
6.0~mg YbRh$_2$Si$_2$ single crystal mounted on the magnetometer,
can be detected with a resolution as high as $10^{-5}$~emu. The
contribution of the sample to the total magnetization of the sample
and pressure cell is larger than 63\% in the entire field and
temperature range. The pressure is determined by the difference
between the superconducting transitions of two small Sn samples; one
placed inside the pressure-transmitting medium (Daphne oil) together
with the YbRh$_2$Si$_2$ sample, the other one outside the pressure
cell. The $T_c$ values are determined using a commercial SQUID
magnetometer. Magnetization data up to 11.5~T to study the
suppression of HF behavior at high fields under pressure have been
published before~\cite{tokiwa-prl05}. Here, we concentrate on the
low-field range close to the AF QCP.

\begin{figure}[tb]
\begin{center}
\vspace{0.5cm}
\includegraphics[width=\linewidth,keepaspectratio]{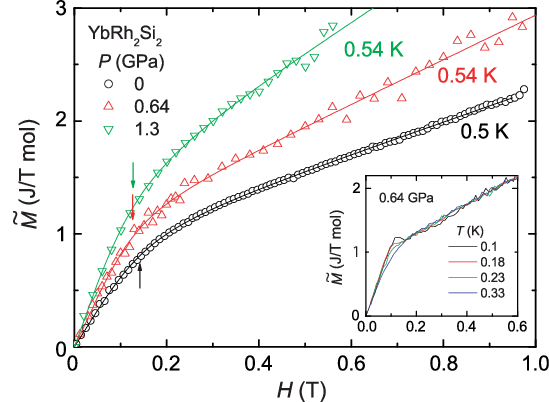}
\end{center}
\caption{(Color online) $\widetilde{M}(H)=M+\chi H$ of YbRh$_2$Si$_2$ at ambient
pressure, as well as 0.64 and 1.28\,GPa for 0.5, 0.54 and 0.54\,K,
respectively. The lines are fits with an empirical crossover
function using the parameters given in Tab.~1. Arrows indicate the
crossover field $H_0$. Inset: $M+\chi H$ under a pressure of
0.64\,GPa at different temperatures.}\label{f2}
\end{figure}

Figure~1 shows the magnetization for $H\perp c$ at 0, 0.64 and
1.28\,GPa at various temperatures. For each temperature, the
magnetization at any field increases significantly under pressure.
The AF critical fields $H_{\rm N}$ indicated by arrows are shifted
from 0.06 to 0.14 and 0.29\,T at 0.64 and 1.28\,GPa, respectively.
At 1.28\,GPa, $M(H)$ shows an additional anomaly at 0.08\,T, at a
second AF transition, which has also been observed in
Yb(Rh$_{0.93}$Co$_{0.07}$)$_2$Si$_2$~\cite{friedemann}. In $M(H)$,
the signature of $H^{\star}$ is a change in slope from higher to
lower values below and above $H^{\star}$, giving rise to a convex
shape of the magnetization curve which broadens with increasing
temperature~\cite{gegenwart-science}. Similar behavior is found
under hydrostatic pressure: As shown in the inset in Fig~1(a), the
$M(H)$ traces at different pressure, scaled by the magnetization
values at 1~T are almost identical at similar temperatures.
Previously, $\widetilde{M}(H)=M+\chi H$ rather than $M(H)$ has been
analyzed for YbRh$_2$Si$_2$ at ambient
pressure~\cite{gegenwart-science} as well as for
Yb(Rh$_{0.93}$Co$_{0.07}$)$_2$Si$_2$~\cite{friedemann}. 
Here $\chi=dM/dH$ denotes the differential susceptibility and
$\widetilde{M}=d(MH)/dH$ is proportional to the field-derivative of
the magnetic free energy contribution
$-MH$~\cite{gegenwart-science}.


\begin{table}[hbp]
    \centering
        \begin{tabular}{c c c c c}
        \hline\hline
        P(GPa) & $T$(K) & $H_{\rm 0}$(T) & $p$ & FWHM(T)\\
        \hline
        0 & 0.5 & 0.14$\pm$0.015 & 3.1$\pm$0.15 & 0.14$\pm$0.04\\
        0.64 & 0.54 & 0.12$\pm$0.015 & 3.4$\pm$0.7 & 0.12$\pm$0.03\\
        1.28 & 0.54 & 0.12$\pm$0.015 & 2.8$\pm$0.7 & 0.125$\pm$0.04\\
        Co-7\%~\cite{friedemann} & 0.5 & 0.125$\pm$0.015 & 2.3$\pm$0.3 & 0.13$\pm$0.04\\
        \hline
        \end{tabular}
        \caption{Parameters for the description of $\widetilde{M}(H)$ shown in Fig.~2 and the respective
        ones for Yb(Rh$_{\rm 0.93}$Co$_{\rm 0.07}$)$_2$Si$_2$ by the empirical crossover function $\int{fdH}$ with $f(H,T)=A_2-(A_2-A_1)/[1+(H/H_0)^p]$~\cite{paschen}.
         Here $A_1$ and $A_2$ represent the slopes of $\widetilde{M}(H)$ below and above the crossover, respectively. $H_0$ is the crossover
         field and the exponent $p$ is a measure for the broadening of the crossover.}
\end{table}

\begin{figure}[tb]
\begin{center}
\vspace{0.5cm}
\includegraphics[width=\linewidth,keepaspectratio]{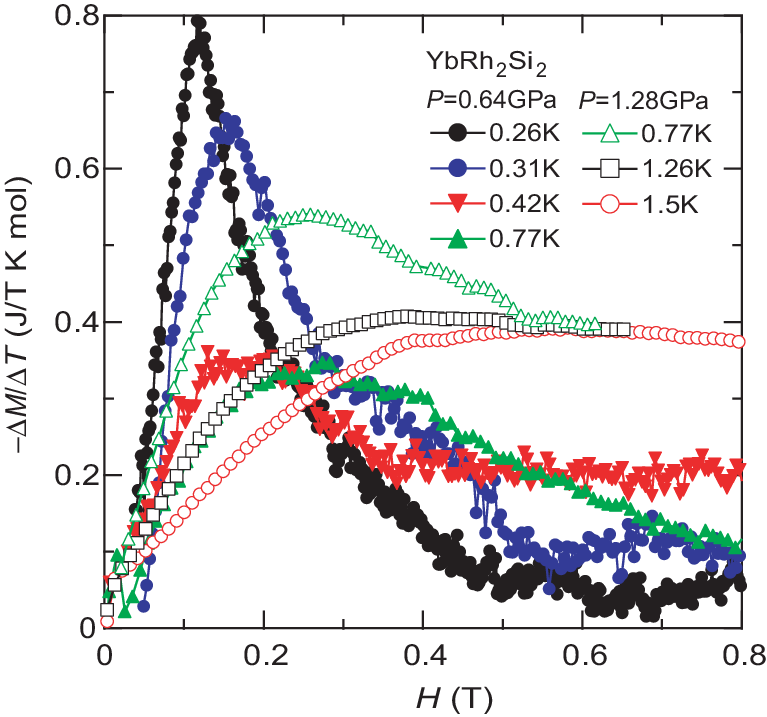}
\end{center}
\caption{(Color online) Magnetization difference divided by temperature increment,
$-\Delta M/\Delta T$, vs magnetic field for YbRh$_2$Si$_2$ under
hydrostatic pressures of 0.64 and 1.28\,GPa, obtained
from isothermal magnetization measurements at different temperatures
(see text).}\label{f3}
\end{figure}

Representative $\widetilde{M}(H)$ data at 0.54\,K (0.50\,K only for
$P=0$) at different pressures are shown in Fig.~2. We used the field integral of the same empirical function of
the form $f(H,T)=A_2-(A_2-A_1)/[1+(H/H_0)^p]$ as used in refs.~\citenum{gegenwart-science,friedemann} to
describe the data with the parameters given in Table 1. We note,
that the obtained position of $H_0$ remains almost unchanged under
pressure. At temperatures below $T_{\rm N}$, the anomaly due to AF
order interferes with the broad kink which marks $T^{\star}$. Under
a pressure of 0.64\,GPa, $\widetilde{M}(H)$ at 0.1\,K exhibits a
peak at $H_{\rm N}$, which shifts to lower field as temperature is
increased up to $T_{\rm N}$, following the AF phase boundary (see
inset). This peak due to AF order disappears at higher temperatures
where only the broad kink remains. This kink indicates the
cross-over field $H_0$ and its position shifts to higher fields as
temperature is increased. Remarkably, both $H_0$ and the full width
at half maximum (FWMH) determined at the various pressures are
almost identical. Also, they nicely agree with the values obtained
for Yb(Rh$_{0.93}$Co$_{0.07}$)$_2$Si$_2$~\cite{friedemann}.

\begin{figure}[t]
\begin{center}
\vspace{0.5cm}
\includegraphics[width=\linewidth,keepaspectratio]{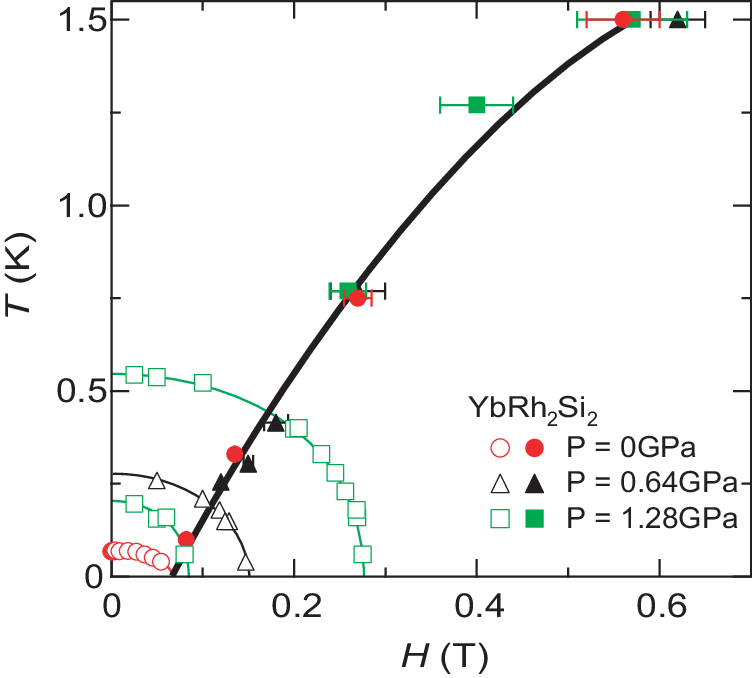}
\end{center}
\caption{(Color online) $H$-$T$ phase diagram of YbRh$_2$Si$_2$ at hydrostatic
pressures of 0, 0.64 and 1.28\,GPa. Open and filled symbols
represent the AF transition and maxima in $-\Delta
M/\Delta T$, respectively. Solid lines are guides to the
eye.}\label{f4}
\end{figure}

We obtained the magnetization difference divided by temperature
increment, $-\Delta M/\Delta T=-\{M(T+\Delta T,H)-M(T-\Delta
T,H)\}/2\Delta T$, from isothermal magnetization data at various
different temperatures under hydrostatic pressure. As discussed
previously~\cite{tokiwa-prl09}, $\Delta M/\Delta T$ probes the field
derivative of the entropy through the Maxwell relation, $\Delta
M/\Delta T\approx dM/dT=dS/dH$. In Fig.~3, we show $-\Delta M/\Delta
T$ at temperatures above $T_{\rm N}$, i.e., outside the AF phase,
since the anomaly due to the AF ordering formation interferes with
the signature of $H^{\star}$. At ambient pressure, $-\Delta M/\Delta
T$ exhibits a maximum very close to $H^{\star}$~(ref.
\citenum{tokiwa-prl09}). Similar maxima, which broaden and shift to
higher fields upon increasing temperature, are also found under
hydrostatic pressure. The positions of the maxima, plotted in
Fig.~4, indicate inflection points in the field dependence of the
entropy, which, as found previously~\cite{tokiwa-prl09}
are related to $H^{\star}$.

As shown in the phase diagram of Fig.~4, the AF phase boundary is
expanded with hydrostatic pressure, and at $P=1.28$\,GPa there are
two AF phases at $T_{\rm N}$ and $T_{\rm L}$~\cite{mederle}. On the
other hand, $T^{\star}$ is independent of pressure, resulting in a
crossing of $T^{\star}(H)$ and $T_{\rm N}(H)$ and different critical
fields where the two temperature scales vanish. This is very similar
to what has been observed for chemically pressurized
Yb(Rh$_{1-x}$Co$_x$)$_2$Si$_2$~\cite{friedemann}. Therefore, the
disorder introduced by partial Co substitution of Rh atoms in the
latter series, which enhances the residual resistivity from
$1\mu\Omega$cm at $x=0$ to $10.7\mu\Omega$cm at $x=0.07$, does not
influence the low-$T$ phase diagram and can not be responsible for
the detachment of the two energy scales. Future pressure experiments
should focus on the nature of the AF QCP. Since under pressure
$H_{\rm N}>H^\star$ it is presumably of standard SDW type. We also
note, that volume expanded Yb(Rh$_{0.94}$Ir$_{0.06}$)$_2$Si$_2$,
with a residual resistivity of $\rho_0=14\mu\Omega$cm, has been
studied by electrical resistivity~\cite{macovei}. Translating the
chemically-induced volume expansion to a pressure of $\Delta
p=-0.06$~GPa has revealed the identical temperature versus pressure
phase diagram (at $H=0$) as found in undoped
YbRh$_2$Si$_2$\cite{mederle}, proving that disorder introduced by
substitution has only minor effects.

In conclusion, we have measured the magnetization of YbRh$_2$Si$_2$
under hydrostatic pressure in order to investigate the boundary of
the AF phase as well as the location of the crossover line
$T^\star(H)$. While the former is increasing with pressure, the
latter remains unchanged, resulting in an intersection between
$T^{\star}(H)$ and $T_{\rm N}(H)$. The entropic signature of
$T^{\star}(H)$ found at ambient pressure~\cite{tokiwa-prl09} is
confirmed. Our results indicate, that the separation of the AF from
the Kondo breakdown QCP in doped YbRh$_2$Si$_2$~\cite{friedemann}
cannot be due to disorder but is clearly related to the change of
the unit-cell volume.

We would like to thank R. Borth and C. Klausnitzer for technical
support and M. Brando, J. Ferstl, S. Friedemann, C. Krellner, T.
Nakanishi, M. Nicklas, Q. Si, G. Sparn and T. Westerkamp for useful
discussions. This work was supported by the DFG research unit 960
"Quantum phase transitions".

\end{document}